\documentclass{PoS}

\usepackage{graphicx}
\usepackage{url}
\usepackage{amsmath}
\usepackage{aas_macros}
\usepackage[table]{xcolor}

\title{VERITAS Observations of High-Mass X-Ray Binary SS 433 }

\ShortTitle{VERITAS Observations of SS 433}
\author{\speaker{Payel Kar} {for the VERITAS Collaboration}\thanks{https://veritas.sao.arizona.edu}\\
        University of Utah\\
        E-mail: \email{payel.kar@utah.edu}}

\abstract{Despite decades of observations across all wavebands and dedicated theoretical modelling, the SS 433 system still poses many questions, especially in the high-energy range. SS 433 is a high-mass X-ray binary at a distance of $\sim 5.5$ kpc, with a stellar mass black hole in a $13$ day orbit around a supergiant $\sim$A7Ib star. SS 433 is unusual because it contains dual relativistic jets with evidence of high-energy hadronic particles. X-ray emission is seen from the central source as well as the jet termination regions, where the eastern and western jets interact with the surrounding interstellar medium. Very-high-energy gamma-ray emission is predicted both from the central source and multiple smaller regions in the jets. This emission could be detectable by current generation imaging atmospheric-Cherenkov telescopes like VERITAS. VERITAS has observed the extended region around SS 433 for $\sim 70$ hours during 2009-2012. No significant emission was detected either from the location of the black hole or the jet termination regions. We report 99\% confidence level flux upper limits above 600 GeV for these regions in the range $(1-10) \times 10^{-13} \mathrm{cm^{-2} \ s^{-1}}$. A phase resolved analysis also does not reveal any significant emission from the extended SS 433 region.}

\FullConference{35th International Cosmic Ray Conference --- ICRC2017\\
		10--20 July, 2017\\
		Bexco, Busan, Korea}

\begin{document}

\section{Introduction}
Object No. 433 in the H-alpha-emission objects catalog by \cite{1977ApJS...33..459S} was recorded exhibiting highly variable radio emissions for the first time in 1978. Since then the SS 433 system has been extensively observed across the electromagnetic spectrum. SS 433 is a binary system located at about $5.5\pm0.2$ kpc from Earth with an accreting black hole of $9\mathrm{M_{\odot}}$ orbitting a $30\mathrm{M_\odot}$ A type supergiant Wolf-Rayet star with an orbital period of 13.082 days \cite{2004ApJ...616L.159B,2002sf2a.conf..317F}. The relativistic, bipolar, hadronic jets spew material with velocity of 0.26c and precess with a period of 162.5 days in cones of half-opening angle $\theta \approx 20^\circ$\cite{2005A&A...437..561C}. The jets are inclined at an angle of $\sim79^\circ$ with the line of sight. The high kinetic luminosity of the jets $L_k\sim10^{39} \mathrm{erg \ s^{-1}}$ \cite{1998AJ....116.1842D} indicates exceptional power output, possibly contributing a reasonable fraction of the Galactic cosmic-ray flux\cite{2002A&A...390..751H}. Models had suggested the cosmic microwave background (CMB) photons could be inverse-Compton upscattered by electrons at the interaction region between the eastern jet and interstellar medium thereby producing TeV gamma rays \cite{1998NewAR..42..579A}. Another model by \cite{2010IJMPD..19..749B} suggests that the interaction region between the western jet and W50 nebula serves as a possible site for GeV-TeV emission arising from relativistic Bremsstrahlung emission. A detection of gamma rays in either eastern or western jet interaction region would provides a unique laboratory to study relativistic shock acceleration with well-known input parameters in addition to confirming the first TeV gamma-ray binary with a microquasar.

\section{VERITAS Observations and Analysis}

The Very Energetic Radiation Imaging Telescope Array Sytem (VERITAS) is an array 4 telescopes that uses the imaging atmospheric-Cherenkov technique to observe gamma rays in the 85 GeV to $>30$ TeV energy range. The telescopes, located at the Fred Lawrence Whipple Observatory (FLWO) in southern Arizona ($31^{\circ}40$' N, $110^{\circ}57$' W, 1.3 km a.s.l.), each have a 12-m diameter reflector focusing light onto a 499 pixel photomultiplier tube (PMT) camera, giving it a $3.5^{\circ}$ field of view. For detailed description and characterization of the instrument refer to \cite{2015ICRC...34..771P}.

The VERITAS observations of SS 433 used in this work were taken between September 2009 and July 2012. This data corresponds to the configuration of the instrument which is after the relocation of one the telescopes to make the array more symmetric and prior to the upgrade of the PMTs in the camera. For a detailed characterization of the instrument during this epoch see \cite{2010HEAD...11.3904P}. Located in the galactic plane, SS 433 is only a few degrees away from the extended TeV source MGRO J1908+06 (R.A. 19h 07m 54s Dec. $+06^{\circ}16'07''$) \cite{2014ApJ...787..166A}. To maximise exposure on both of these objects, the camera was pointed at a central location between them, thereby accommodating both SS 433 and MGRO J1908+06 within the field of view. A total of $\sim70$ hours of quality-selected data (live time) were obtained, all with the SS 433 central black hole within $1.5^{\circ}$ from the camera center. Table \ref{veritasobs} shows a summary of the VERITAS observations of the SS 433 region. 

\begin{table}[h]
\centering
    \caption{VERITAS observations of SS 433}% 
	\label{veritasobs}
    \vspace{\baselineskip}
    \begin{center}
        \begin{tabular}{|c | c | c | c | c |}
            \hline \hline 
			Year & Exposure [h] & No. of Runs & Camera Offset $(^{\circ})$ & Mean Elevation $(^{\circ})$\\ 
      	  	\hline 
		    2009 & 9.33 	& 27	& 0.7-1.0 	& 59 	\\
			2010 & 10.5 	& 37	& 0.7-0.8 	& 60 	\\
			2011 & 26.11 	& 88	& 0.05-1.1	& 61.2 	\\
			2012 & 25.23	& 79	& 0.7-1.5	& 62.4	\\
            \hline 
			Total & 71.17	& 231 & 0.05-1.5	& 61	\\
			\hline \hline
        \end{tabular}
    \end{center}
\end{table}

The SS 433 system is extended in the sky over a $2^{\circ}\times2^{\circ}$ region. Four locations were selected as regions of interest (ROIs) for this work. These are the central location of the black hole, two locations e1 and e2 in the eastern jet interaction region and, one location w2 in the western jet interaction region. These ROIs were selected from a previous X-ray spectral analysis study using ROSAT/ASCA data, as they were deemed possible sites for VHE emission (see \cite{1997ApJ...483..868S,2009AA...497..325B}). 

Each of the pre-selected ROIs were searched for point source emission. The data was analyzed using a standard analysis package, implementing a specialized Boosted Decision Tree (BDT) technique \cite{2017ICRC...M,2017APh....89....1K}. According to the model outlined in \cite{2008MNRAS.387.1745R}, the precessing jet should cause phase-based flux variations from SS 433. Setting $\phi=0$ at JD 2443507.47 when the accretion disc is maximally open\footnote{In this orientation of the disc the gamma rays emitted in the innermost regions would escape without travelling through the thick extended and disc undergoing $\gamma\gamma$ absorption} to the observer and using a precession period of 162.5 days, the data are divided into five phase bins of width $\Delta\phi=0.2$ \cite{2013MNRAS.436.2004C}. Each ROI is also searched for TeV emission during each of the five individual phases. 

\section{Results}
\begin{figure}[t]
\centerline{\includegraphics[scale = 0.5]{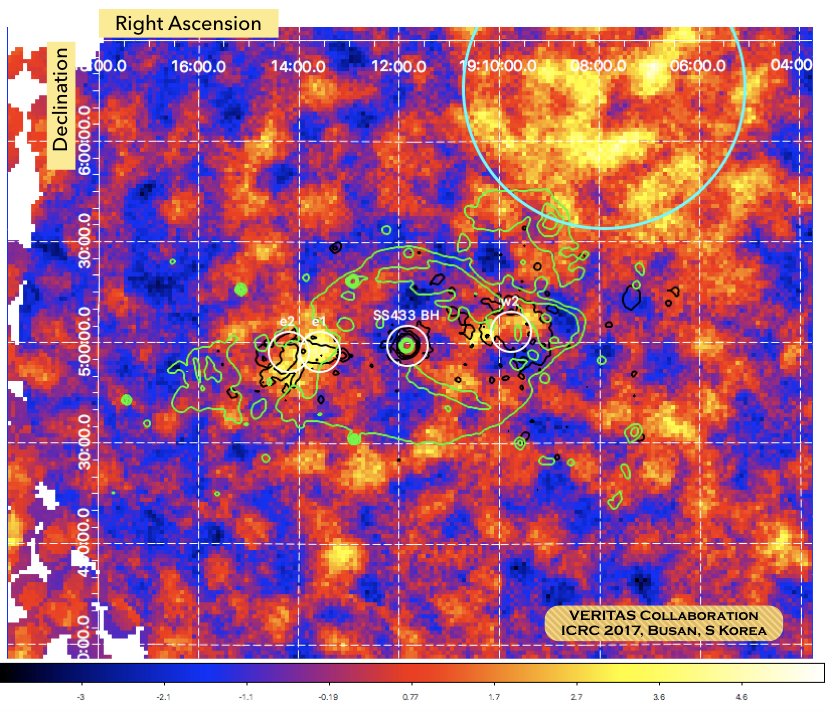}}
\caption{ VERITAS significance skymap of the SS 433 region. The 4 regions of interest, SS 433 black hole, w2, e1 and e2 are marked in white. Contours in green are from radio observations \cite{1998AJ....116.1842D}. Contours in black are X-ray observations from ROSAT/ASCA \cite{1997ApJ...483..868S}. Bright extended emission from MGRO J1908+06 is seen on the top right of the skymap located within the cyan circle \cite{2014ApJ...787..166A}, this region is excluded from the background estimation of the analysis.}%
\label{ss433_skymap}
\end{figure}

No significant emission is found from the location of the black hole or any of the three jet and interstellar medium interaction regions w2, e1 or e2. The VERITAS skymap of the SS 433 region is shown in Figure \ref{ss433_skymap}. The location of the ROIs are marked in white. The skymap is overlayed with X-ray emission contours in black from ROSAT/ASCA \cite{1997ApJ...483..868S}. The green radio contours from VLA \cite{1998AJ....116.1842D} are also shown on the VERITAS skymap. There is weak evidence of emission ($\sim4\sigma$ above the background not accounting for the number of statistical trials conducted in this analysis) from e1, one of the ROIs in the eastern jet interaction region, but as the significance is below $5\sigma$, no positive detection is claimed.

\begin{figure}[t]
\centerline{\includegraphics[scale = 0.4]{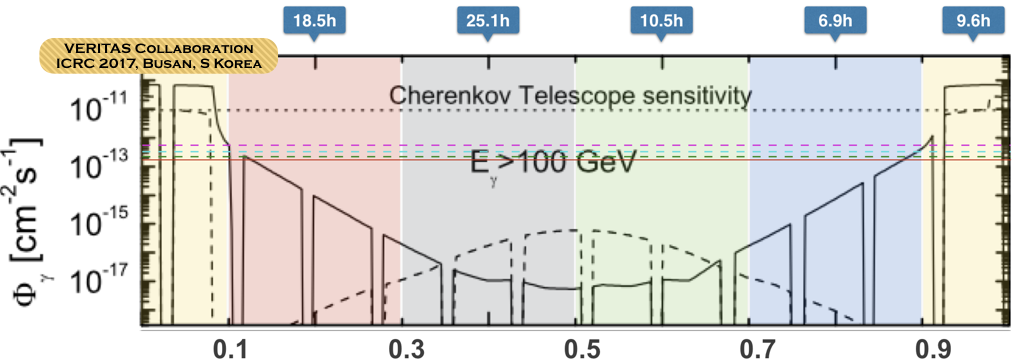}}
\caption{ VERITAS 99\% confidence level flux upper limits $>600$ GeV for the 4 regions of interest, SS 433 black hole in red, w2 in green, e1 in magenta and e2 in cyan. These upper limits are overlayed on a predicted model of phased emission adapted from \cite{2008MNRAS.387.1745R}. The doted line shows prediction from 2005 of current generation Cherenkov Telescope sensitivity. The modelled contribution to gamma-ray flux $>100$ GeV from the two jets are shown separately, where the solid line represents the approaching or western jet and the dotted line represents the recessing or eastern jet}%
\label{reynoso_ul}
\end{figure}

A plot of the model for flux variation due to the jets precession adapted from \cite{2008MNRAS.387.1745R} is overlaid with integral flux upper limits from the four ROIs, and shown with respect to the phase-based flux variation model in Figure \ref{reynoso_ul}. For each of the phase bins, 99\% confidence level flux upper limits above 600 GeV are also calculated for the four ROIs, a summary of all the upper limits is also presented in Table \ref{ultable}. 

% \begin{figure}[b]
% \centerline{\includegraphics[scale = 0.3]{phase_ul.png}}
% \caption{ Phase based VERITAS 99\% confidence level flux upper limits $>600$ GeV for the 4 regions of interest in the SS 433 system, SS 433 black hole in red, w2 in green, e1 in magenta and e2 in cyan.}%
% \label{phase_ul}
% \end{figure}

\section{Discussion}

Analysis of 10.4 hours of data recorded between September 2007 and July 2008 which is not included in this analysis, found a $4.9\sigma$ excess at w2 (not accounting for statistical trials), a location in the interaction region between the western jet and the surrounding interstellar medium \cite{2010PhDT.......228G}. This motivated further observations by VERITAS during the 2009-2012 period. The additional observations presented in this work did not detect any statistically significant VHE emission from w2 or any of the other selected regions in the SS 433 system, namely the position of the black hole and the two locations e1 and e2, which are situated in the interaction region of the eastern jet and the surrounding interstellar medium. Out of the total $\sim70$ h of VERITAS data, nearly 26 h are taken in an unfavorable phase when the star obscures the disc and jet. In this orientation the high density of matter from the star and the surrounding W50 nebula subjects gamma rays to a high degree of absorption by mechanisms like $\gamma\gamma$ interactions with ambient soft photons and by $\gamma N$ (where $N$ represents nucleons) interactions with disc and stellar matter. For details of various possible absorption mechanisms see \cite{2008APh....28..565R}.

X-ray emission detected from the inner and outer lobes of SS 433 may be of non-thermal origin as suggested in \cite{1994PASJ...46L.109Y,1997ApJ...483..868S}. Recently, gamma-ray emission from the direction of SS 433 was detected by \textit{Fermi}-LAT revealing a very peculiar spectral energy distribution of the source with a distinct maximum at 250 MeV and extending only up to 800 MeV \cite{2015ApJ...807L...8B}. Although it is not yet clear whether the detected emission is associated with SS 433 the lack of other plausible counterparts in the region supports this hypothesis. If this is the case, the cutoff in the GeV spectrum implies that the maximum energies to which electrons and protons are accelerated are just a few GeV \cite{2015ApJ...807L...8B}. Thus, TeV emission from the source is not expected. However, such low maximum energies of accelerated particles are in contradiction with the non-thermal interpretation of the X-ray emission which requires electrons with energies in the order of 10-100 TeV \cite{1994PASJ...46L.109Y,1997ApJ...483..868S}. It is also unclear where the particles responsible for the HE gamma-ray emission are accelerated. The poor angular resolution of \textit{Fermi}-LAT at these energies ($\geq1.5^{\circ}$ at energies of about 300 MeV) \cite{2015ApJ...807L...8B} does not allow it to resolve the origin of the emission. However, the lack of flux variability suggests that the emission is generated in outer regions far from the binary system, since otherwise it would be subject to strong phase dependent absorption by photo-hadronic interactions with disk and stellar matter, which would show evidence of precessional modulation \cite{2008APh....28..565R}. If this is the case, possible, detectable very-high-energy emission from inner regions of jets cannot be ruled out.
%%%%%%%%%%%%%%% SS 433 BH position%%%%%%%%%%%%%
\begin{table}[h]
	\center
	\caption{VERITAS upper limits of the SS 433 binary system }
	\label{ultable}
	\vspace{\baselineskip}
\begin{tabular}{ |c|c|c|c|c|c|  }
 \hline \hline & & & & & 99\% Flux UL \footnotemark\\
 Phase & $N_{ON}$ & $N_{OFF}$ & $\sigma$ & Live time (min) & ($>600$ GeV)\\ & & & & & $[ \mathrm{cm^{-2} \ s^{-1}}]$\\
 \hline
 \multicolumn{6}{|c|}{\cellcolor{red!50}Upper limits of the SS 433 Black hole position} \\
 \hline
 0.1-0.3 & 52 & 586 & 0.2 & 1113.7 & 4.12$\times 10^{-13}$\\
 0.3-0.5 & 62 & 731 & -0.1 & 1515.9 & 3.39$\times 10^{-13}$ \\
 0.5-0.7 & 35 & 357 & 0.7 & 630.5 & 7.95$\times 10^{-13}$ \\
 0.7-0.9 & 22 & 231 & 0.3 & 414.8& 7.90$\times 10^{-13}$\\
 0.9-1.1 & 25 & 282 & 0.0 & 577.1 & 5.70$\times 10^{-13}$\\
 \hline
 All & 196 & 2195 & 0.4 & 4272.1 & 2.29$\times 10^{-13}$ \\
 \hline
 \multicolumn{6}{|c|}{\cellcolor{green!50}Upper limits of western lobe (w2)} \\
 0.1-0.3 & 57 & 821 & 0.6 & 1113.7 & 4.36$\times 10^{-13}$ \\
 0.3-0.5 & 69 & 1080 & 0.0 & 1515.9 & 3.16$\times 10^{-13}$\\
 0.5-0.7 & 42 & 546 & 1.1 & 630.5 & 8.07$\times 10^{-13}$ \\
 0.7-0.9 & 31 & 325 & 2.0 & 414.8 & 1.17$\times 10^{-12}$\\
 0.9-1.1 & 26 & 463 & -0.6 & 577.1 & 4.05$\times 10^{-13}$\\
 \hline
 All & 225 & 3245 & 1.2 & 4272.1 & 2.66$\times 10^{-13}$\\
 \hline 
 \multicolumn{6}{|c|}{\cellcolor{magenta!50}Upper limits of eastern lobe (e1)} \\
 0.1-0.3 & 49 & 563 & 1.1 & 1113.7 & 6.22$\times 10^{-13}$\\
 0.3-0.5 & 68 & 583 & 2.6 & 1505.8 & 8.54$\times 10^{-13}$\\
 0.5-0.7 & 36 & 278 & 2.5 & 630.5 & 1.45$\times 10^{-12}$\\
 0.7-0.9 & 23 & 218 & 1.6 & 414.8 & 1.40$\times 10^{-12}$\\
 0.9-1.1 & 23 & 280 & 0.5 & 577.1& 7.97$\times 10^{-13}$\\
 \hline
 All & 199 & 1925 & 3.7 & 4262.1& 5.68$\times 10^{-13}$ \\
 \hline
 \multicolumn{6}{|c|}{\cellcolor{cyan!50}Upper limits of eastern lobe (e2)} \\
 \hline
 0.1-0.3 & 39 & 427 & -0.1 & 1113.7 & 4.47$\times 10^{-13}$ \\
 0.3-0.5 & 61 & 435 & 3.0 & 1445.7 & 1.03$\times 10^{-12}$\\
 0.5-0.7 & 21 & 231 & 0.2 & 630.5 & 7.79$\times 10^{-13}$\\
 0.7-0.9 & 15 & 172 & -0.1 & 414.8 & 9.23$\times 10^{-13}$\\
 0.9-1.1 & 17 & 172 & 0.4 & 577.1 & 7.77$\times 10^{-13}$\\
 \hline
 All & 153 & 1442 & 1.8 & 4202.0 & 4.21$\times 10^{-13}$\\
 \hline \hline
\end{tabular}
\end{table} 

\section{Conclusion}
A targeted point source search was applied to various locations in the SS 433 system. VERITAS did not detect any significant TeV emission from these predefined regions which could imply that SS 433 may be not be intrinsically as luminous above 600 GeV as was previously thought or if gamma rays are produced in the inner regions of the system, they are absorbed significantly by the surrounding W50 nebula. The possibility of extended emission from the jet and interstellar medium interaction regions, as suggested by the $\sim4\sigma$ excess from e1 region in the eastern jet, and a study exploring the skymap with extended source analysis methods is currently being done. \\

\noindent\textbf{Acknowledgments} \\ \footnotetext{All  99\% confidence level upper limits calculated using Rolke metod \cite{2001NIMPA.458..745R}}
The author would like to thank Gloria Dubner and Samar Safi-Harb for providing the radio and X-ray contours. This research is supported by grants from the U.S. Department of Energy Office of Science, the U.S. National Science Foundation and the Smithsonian Institution, and by NSERC in Canada. We acknowledge the excellent work of the technical support staff at the Fred Lawrence Whipple Observatory and at the collaborating institutions in the construction and operation of the instrument. 

\bibliography{skeleton}
%apjmnemonic
% \begin{thebibliography}{99}
% \bibitem{...}
% ....
%
% \end{thebibliography}

\end{document}